\documentclass[twocolumn,twoside,slac_two]{revtex4}
\usepackage{graphicx}
\usepackage{fancyhdr}
\begin{document}

\title{Observations of Gamma-ray Bursts with  ASTRO-H and  Fermi}

%

\author{M. Ohno$^1$, T. Kawano$^1$, M. S. Tashiro$^2$, H. Ueno$^2$, D. Yonetoku$^3$, H. Sameshima$^4$, T. Takahashi$^4$, H. Seta$^5$, R. Mushotzky$^6$, K. Yamaoka$^7$, {\it ASTRO-H} SWG team and {\it Fermi} LAT/GBM collaborations}
%



\affiliation{$^1$Department of Physical Sciences, Hiroshima University, 1-3-1 Kagamiyama, Higashi-Hiroshima, Hiroshima 739-8526}
\affiliation{$^2$Department of Physics, Saitama University, 255 Shimo-Okubo, Sakura-ku, Saitama-shi, Saitama 338-8570}
\affiliation{$^3$Department of Physics, Kanazawa University, Kadoma-cho, Kanazawa, Ishikawa 920-1192 }
\affiliation{$^4$Institute of Space and Astronautical Science Aerospace Exploration Agency (ISAS/JAXA), 3-1-1 Yoshinodai, Sagamihara, Kanagawa 229-8510 }
\affiliation{$^5$Department of Physics, Rikkyo University, Nishi-Ikebukuro, Toshimaku, Tokyo 171-8501}
\affiliation{$^6$Department of Astronomy, University of Maryland College Park, MD 20742-2421}
\affiliation{$^7$Solar-Terrestrial Environment Laboratory, Nagoya University, Furo-cho, Chikusa-ku, Nagoya 464-8601}

\begin{abstract}

{\it ASTRO-H}, the sixth Japanese X-ray observatory, which is scheduled to be launched by the end of Japanese fiscal year 2015 has a capability to observe the prompt emission
from Gamma-ray Bursts (GRBs) utilizing BGO active shields for the soft gamma-ray detector (SGD). The effective area of the SGD shield
detectors is very large and its data acquisition system is optimized for short transients such as short GRBs. Thus, we expect to perform more
detailed time-resolved spectral analysis with a combination of {\it ASTRO-H} and Fermi LAT/GBM to investigate the gamma-ray emission
mechanism of short GRBs. In addition, the environment of the GRB progenitor should be a remarkable objective from the point of view of the
chemical evolution of high-z universe. If we can maneuver the spacecraft to the GRBs, we can perform a high-resolution spectroscopy of the
X-ray afterglow of GRBs utilizing the onboard micro calorimeter and X-ray CCD camera.

\end{abstract}

\maketitle

\thispagestyle{fancy}


\section{Introduction}

Gamma-ray Bursts (GRBs) are one of the most energetic explosion in the universe, but 
there are still many issues to be understood such as gamma-ray emission mechanism of prompt emission, physical composition of jet outflow, and environment of progenitor. GRBs are also known to be originated at cosmological distance and they would be useful to explore the chemical evolution of 
high-z universe. 

{\it ASTRO-H} is the sixth X-ray observatory from Japan, which is scheduled to be launched by the end of Japanese fiscal year 2015. 
Four onboard instruments of {\it ASTRO-H}, the high-resolution
X-ray micro-calorimeter (Soft X-ray Spectrometer: SXS), X-ray CCD camera (Soft X-ray Imager: SXI), 
Hard X-ray Imager (HXI), and Soft Gamma-ray Detector (SGD) realize wide-band and high-sensitivity
observation from 0.3 to 600 keV energy band. The high-resolution spectroscopy by SXS and X-ray observations
with enough photon statistics by SXI could be very powerful tool to investigate spectral features and detail of X-ray absorption structure in the afterglow spectrum of GRBs. And also, high-sensitive hard X-ray observation by HXI might observe interesting features from afterglow in hard X-rays. In addition to such afterglow observations
by focal plane instruments, {\it ASTRO-H} is also able to observe the prompt gamma-ray emission utilizing SGD. Therefore, {\it ASTRO-H} will bring us a comprehensive observation of GRBs from prompt gamma-ray emission to
subsequent X-ray afterglow emission. In this paper, we demonstrate a capability of GRB observation
by {\it ASTRO-H}.

\section{Prompt emission observation by the SGD shield}

\begin{figure}
\includegraphics[width=65mm]{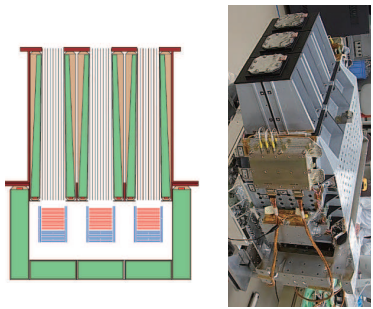}
\caption{A schematic picture and a real flight model picture of the SGD. The three main detectors are located inside of BGO crystals.}
\label{fig:SGDpic}
\end{figure}

Our understanding of gamma-ray emission mechanism of GRBs, especially for short duration GRBs 
is still poor. One of key observation to solve such problem is time-resolved 
spectroscopy as was performed for long duration GRBs. However, photon statistics of short GRBs is too low to 
perform such time-resolved analysis, and therefore, observation of short GRBs with large effective area is important. {\it ASTRO-H} has capability to observe prompt gamma-ray emission of short GRBs
with large effective area and good time-resolution utilizing SGD. The main detector, Compton camera of the SGD is surrounded by large 25 BGOs to reduce background by anti-coincidence technique as shown in Fig \ref{fig:SGDpic}.  
Thanks to its large geometrical area and high gamma-ray stopping power of BGO crystal, the effective area
of those ``shield`` detectors retain $\sim$ 800 cm$^2$ even at 1 MeV. Therefore, the SGD shield detector
acts as a powerful all-sky monitor like Suzaku WAM\cite{Yamaoka2009}. We have developed
the SGD shield detector so that we can observe short transients such as short GRBs or Soft Gamma Repeaters with many advantages compared with Suzaku WAM. Table \ref{tab:WAMspec} shows some specifications of the SGD shield detector as an all-sky monitor comparing with Suzaku WAM.
The main advantage of the SGD shield detector is that it can obtain spectral information with very large
effective area. 
We also improved data acquisition timing of GRB data of the SGD shield so that we can transfer GRB data
to the spacecraft soon ($\sim$10 min) after trigger and we can set the trigger to be ready for the next GRB. This enable us to improve the efficiency of GRB observation.

\begin{table}[t]
\begin{center}
\caption{Performance of the SGD shield detector as all-sky monitor comparing with Suzaku-WAM}
\begin{tabular}{|c|c|c|}
\hline
\hline
 & SGD shield & Suzaku WAM \\
\hline
Time resolution & 16 ms & 16 ms \\
Time coverage & 5.376 s &  64 s \\
 & (-1.376 to 4.0 s) & (-8 to 56 s)\\
Spectral channels & 32 ch & 4 ch \\
Energy range & 150 -- 5000 keV & 50 -- 5000 keV \\
Effective area (1MeV)& $\sim$ 800 cm$^2$ & $\sim$ 400 cm$^2$\\
\hline
\end{tabular}
\label{tab:WAMspec}
\end{center}
\end{table}

\begin{figure}[t]
\includegraphics[width=85mm]{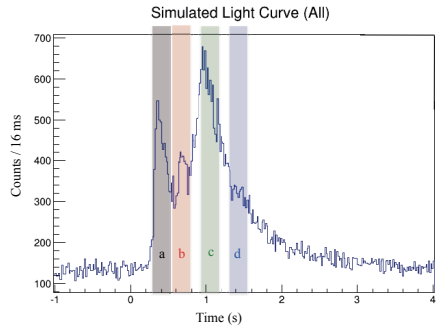}
\caption{An example of light curve simulation of bright short GRB with photon flux of a few tens of photons s${-1}$ cm$^{-2}$ by the SGD shield. Each hatched region show the time window for the demonstration of time-resolved spectral analysis in the below figure.}
\label{fig:LCsim}
\end{figure}
\begin{figure}
\rotatebox{-90}{\includegraphics[width=55mm]{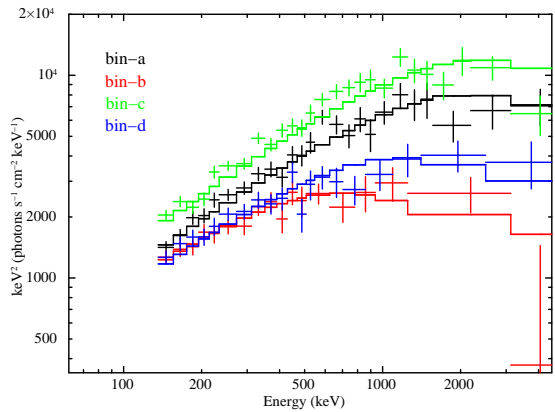}}
\caption{A simulated time resolved spectral analysis using above simulated light curve data.}
\label{fig:Specsim}
\end{figure}

Figure \ref{fig:LCsim}  shows an example of simulated light curve of bright 
short GRB with peak flux of about a few times of 10 photons s$^{-1}$ cm$^{-2}$ in 1 second time scale. In this simulation, we consider poisson fluctuation in each time bin of the SGD shield (16 ms) and we assumed 
 simple Band function with low-energy index $\alpha = -0.8$ and high-energy 
index $\beta = -2.3$.   The peak energy E$_{\rm peak}$ has changed depending on the flux. Figure \ref{fig:Specsim} shows the time-resolved spectrum extracted with 0.1 s time windows, 
which are shown by hatched area in the figure \ref{fig:LCsim}.  
We can see that the simulated light curve exhibit fine time structure and extracted time-resolved 
spectra show clear evolution of E$_{\rm peak}$. Therefore, we can expect to have such GRB data
with the SGD shield.
After launch of {\it ASTRO-H},  GRB data observed by the SGD shield will be publicly available as well as
Suzaku WAM.  The GRB observation by the SGD shield can provide complementary dataset to Fermi-GBM. Based on simultaneously detection rate between Suzaku-WAM and Fermi-GBM, about a half of
GRBs detected by Fermi-GBM are expected to be also detected by the SGD shield.

\section{ToO observations of afterglow with SXS and SXI}

\begin{figure}[t]
\includegraphics[width=85mm]{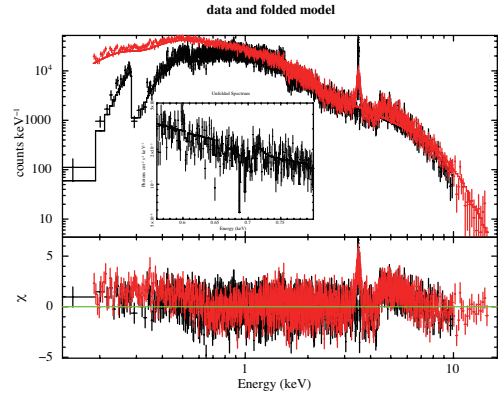}
\caption{{\it ASTRO-H} simulation of GRB afterglow spectrum with warm absorber model (see text in detail).
The enlarged structure around 0.6-0.8 keV, where the most prominent absorption features can be seen are also shown in the inset. Bottom part shows the residuals from single power-law model with absorption from cold materials.}
\label{fig:afterglowsim}
\end{figure}

\begin{figure}
\rotatebox{-90}{\includegraphics[width=65mm]{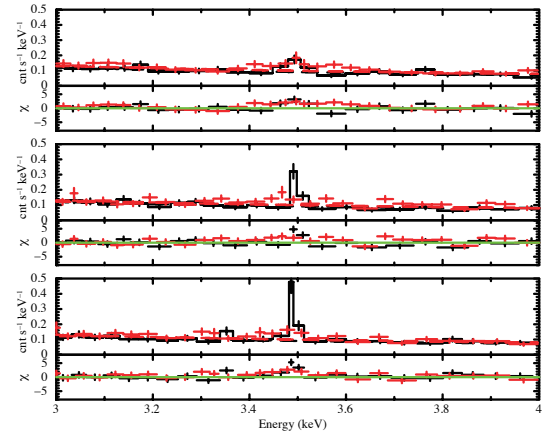}}
\caption{10 ks simulation of around iron-K emission line. Each panel show the simulation with different
intrinsic line width ($\sigma$= 30 eV, 10 eV, and 5 eV from top to bottom). Lower part in each panel shows the residual from single power-law model.}
\label{fig:FeLinesim}
\end{figure}

As for the X-ray afterglow of GRBs, which is widely believed that the X-ray emission is coming from synchrotron
emission due to accelerated electrons in the external forward shock. Therefore, most of X-ray spectrum of
afterglow show featureless simple power-law shape. However, there are several reports of marginal detection of 
spectral features such as iron-K emission line, its recombination edge, and several lines due to light metals
\cite{Piro1999},\cite{Yoshida1999},\cite{Reeves2002}.
Although, they are still controversial  probably because of limited statistics and/or spectral resolution, such spectral features would be very important
to investigate physical conditions of GRB jet and composition of environment of GRB host, and also they are useful to 
 determine the redshift of GRB by X-ray observation itself. 
In addition to such spectral features, Behar et al. (2011) and  Starling et al. (2013)
have pointed out the evidence of excess absorption in soft X-ray energy band using huge sample of Swift X-ray
afterglow observation. One possibility of origin for such excess absorption is contribution of absorption by intergalactic medium (IGM). Therefore, detail spectroscopy in soft X-ray band could give important information
to investigate the property of IGM in high-z universe. An X-ray observation with high spectral resolution is a
key to solve above open questions in GRB afterglow. Those emission line and/or absorption line spectral features can be investigated by unprecedented high energy resolution spectroscopy by 
{\it ASTRO-H} SXS, and detail of continuum structure can be determined by SXI. Figure \ref{fig:afterglowsim} shows a 100 ks {\it ASTRO-H} simulation with SXS and SXI. Here we assumed  GRB 991216  spectrum as the baseline model fro simulation. In this model, 2$-$10 keV flux is set to be $3\times10^{-12}$ erg cm$^{-2}$ s$^{-1}$. This GRB has been reported to have iron-K line and its recombination edge\cite{Piro2000} and thus we include those spectral features in the simulation.  We also added soft X-ray lines reported by Reeves et al. (2003) for GRB 011211, and intergalactic warm absorbers (WHIM) with the temperature of 10$^5$ K, the column density of N$_{\rm H}$ of 10$^{22}$ cm$^{-2}$, and we put those 
absorption material on redshift of z=0.1.  From this simulation, we can see that the iron-K related spectral features can be detected clearly by {\it ASTRO-H} if they are really exist. In addition, some resonance absorption lines due to WHIM are also detectable with about 4 sigma significance level, thanks to high energy resolution of SXS. Figure \ref{fig:FeLinesim} shows the same simulation with
figure \ref{fig:afterglowsim} but with shorter exposure of 10 ks and we changed intrinsic line width from
5 eV to 30 eV. We can clearly detect the iron-K line emission if it is intrinsically narrow with $\sigma < 10$ eV with short exposure of 10 ks. This indicates that we can investigate the time variability of such
iron-K line emission, which is useful to discuss the environment of host galaxy o GRBs.


\section{Expected event rate of ToO observations of GRB afterglow with {\it ASTRO-H}}

\begin{figure}[t]
\rotatebox{-90}{\includegraphics[width=55mm]{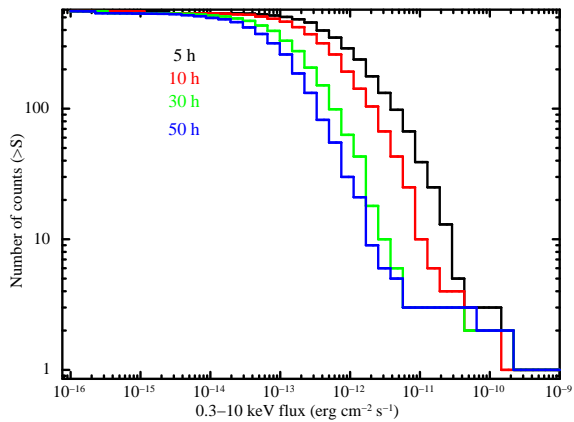}}
\caption{Estimated luminosity functions of GRB afterglow based on 6-yeas Swift-XRT data base. Different colors show the function corresponds to the different observation start time from GRB trigger.
(black: 5 hours, red: 10 hours, green: 30 hours, and blue: 50 hours).}
\label{fig:GRBlumin}
\end{figure}

As we shown in previous section, {\it ASTRO-H} has a capability of detection of spectral features
from GRB afterglow such as iron-K emission line and resonance absorption lines due to intergalactic
warm absorbers. Then, we have to estimate how many number of GRBs we can observe with such
interesting spectral features by {\it ASTRO-H}. For this purpose, we calculated a luminosity function of
GRB afterglow based on 572 samples of 6-years Swift-XRT data base which is publicly available in the web 
page \cite{Evans2009}. Figure \ref{fig:GRBlumin} shows the luminosity functions of GRB afterglow for
several times after GRB trigger. From this result, we can see that about 10 GRBs/year are expected 
which have 10$^{-12}$ erg s$^{-1}$ cm$^{-2}$ flux level, which corresponds to that of we used in the iron-K line simulation in Fig \ref{fig:FeLinesim}, even 30 hours after the trigger. This means
that if we can slew the {\it ASTRO-H} spacecraft within 1-day after the trigger, we could have 10 GRBs/year
samples for possible iron-K line search with {\it ASTRO-H}.

\section{Summary}

In this paper, we demonstrated the capability of GRB observation by {\it ASTRO-H}. As for the prompt gamma-ray emission, the SGD shield detector will act as powerful GRB monitor with very large 
effective area. Especially for the short GRBs, time-resolved spectroscopy with good photon statistics
can be performed by the SGD shield and such GRB data can be a complimentary data set to Fermi-GBM. About a half of GRBs that are detected by Fermi-GBM are also expected to be observed by the
the SGD shield simultaneously. High resolution spectroscopy by {\it ASTRO-H} SXS and SXI is expected to
reveal the existence of spectral features in the GRB afterglow spectrum such as emission lines from
iron-K and/or other light metals, and absorption by intergalactic medium. The expected event rate
of GRBs which can be used for such search of spectral features is estimated to be $\sim$ 10 GRBs/year, if we can slew the spacecraft within 1-day after the GRB trigger. More details about {\it ASTRO-H} 
observation of GRB afterglow can be found in the {\it ASTRO-H} white paper\cite{tashiro14}.

\bigskip 

\begin{thebibliography}{99}   


\bibitem{tashiro14}
M. S. Tashiro, et al. "ASTRO-H White Paper - Chemical Evolution in High-z Universe", arXiv:1412.1179, 2014.

\bibitem{Piro1999}
Piro, L., et al. "The X-Ray Afterglow of the Gamma-Ray Burst of 1997 May 8:Spectral Variability and Possible Evidence of an Iron Line",  ApJ, 514, L73, 1999.

\bibitem{Yoshida1999}
Yoshida, A., et al. , Proc. of ``Gamma-ray bursts in the afterglow era'’ , (A\&AS, 138), 433, 1999.

\bibitem{Piro2000}
Piro, L. et al. "Observation of X-ray Lines from a Gamma-Ray Burst (GRB991216): Evidence of Moving Ejecta from the Progenitor", Science, 290, 955, 2000.

\bibitem{Reeves2002}
Reeves, J. N. et al. "The signature of supernova ejecta in the X-ray afterglow of the γ-ray burst 011211", Nature, 416, 512, 2002.





\bibitem{Reeves2003}
Reeves, J. N. et al. "Soft X-ray emission lines in the afterglow spectrum of GRB 011211: A detailed XMM-Newton analysis", A\&A, 403, 463, 2003.


\bibitem{Evans2009}
Evans, P. A. et al.  "Methods and results of an automatic analysis of a complete sample of Swift-XRT observations of GRBs", MNRAS, 397, 1177, 2009

\bibitem{Yamaoka2009}
Yamaoka, K, et al. "Design and In-Orbit Performance of the Suzaku Wide-Band All-Sky Monitor", PASJ, 61S, 35, 2009.

\bibitem{Behar2011}
Behar, E., Dado, S., Dar, D., Laor, A. "Can the Soft X-Ray Opacity Toward High-redshift Sources Probe the Missing Baryons?",  ApJ, 734, 26, 2011.

\bibitem{Starling2013}
Starling, R. L. C., et al. "X-ray absorption evolution in gamma-ray bursts: intergalactic medium or evolutionary signature of their host galaxies", MNRAS, 431, 3159, 2013.





\end{thebibliography}

\end{document}